\documentclass[aps,prl,twocolumn,showpacs,letterpaper]{revtex4-1}

\usepackage{graphicx}% Include figure files
\usepackage{dcolumn}% Align table columns on decimal point
\usepackage{bm}% bold math
\usepackage{amsmath}
\usepackage{multirow}

\begin{document}

\newcommand{\mevcc}{\!\mathrm{MeV}\!/c^2}
\newcommand{\mevc}{\!\mathrm{MeV}/\!c}
\newcommand{\mev}{\!\mathrm{MeV}}
\newcommand{\gevcc}{\!\mathrm{GeV}/\!c^2}
\newcommand{\gevc}{\!\mathrm{GeV}/\!c}
\newcommand{\gev}{\!\mathrm{GeV}}

\title{Observation of $\bm{\eta_b(2S)}$ in $\bm{\Upsilon(2S)\to\gamma\eta_b(2S)}$, $\bm{\eta_b(2S)\to\mathrm{hadrons}}$, and Confirmation of $\bm{\eta_b(1S)}$}

\author{S.~Dobbs}
\author{Z.~Metreveli}
\author{Kamal~K.~Seth}
\author{A.~Tomaradze}
\author{T.~Xiao}
\affiliation{Northwestern University, Evanston, Illinois 60208, USA}

%\date{July 14, 2009}
\date{\today}

\begin{abstract} 
The data for 9.3 million $\Upsilon(2S)$ and 20.9~million $\Upsilon(1S)$ taken with the CLEO~III detector has been used to study the radiative population of states identified by their decay into twenty~six different exclusive hadronic final states.   In the $\Upsilon(2S)$ decays an enhancement is observed at a $\sim5\sigma$  level at a mass of $9974.6\pm2.3(\mathrm{stat})\pm2.1(\mathrm{syst})$~MeV.  It is attributed to $\eta_b(2S)$, and corresponds to the $\Upsilon(2S)$ hyperfine splitting of $48.7\pm2.3(\mathrm{stat})\pm2.1(\mathrm{syst})$~MeV.
In the $\Upsilon(1S)$ decays, the identification of $\eta_b(1S)$ is confirmed at a $\sim3\sigma$ level with $M(\eta_b(1S))$ in agreement with its known value. 
\end{abstract}

\maketitle

The spin--dependent interaction between constituents of a composite system is of general interest, from hydrogen and positronium to mesons and baryons.
The hyperfine interaction between quarks is one of the most important components of the spin--dependent QCD interaction.  It has been studied for charm quarks by measurements of hyperfine splittings between spin--triplet and spin--singlet states, $M(n^3L)-M(n^1L)$, of 1S, 2S, and 1P states of charmonium~\cite{pdg}.  For the bottom quarks, measurements have been reported recently for the 1S~hyperfine splittings~\cite{etab} and for the 1P and 2P hyperfine splittings~\cite{hb} of bottomonium states.  While $\eta_b(1S)$, the bottomonium ground state, has been identified, the radially excited $\eta_b(2S)$ has not been identified so far, and the hyperfine splitting of the bottomonium 2S states, $\Delta M_{hf}(2S)\equiv\Upsilon(2S)-\eta_b(2S)$, is not known.  In this letter, we report on the observation of $\eta_b(2S)$ in its formation in the radiative decay of $\Upsilon(2S)$ and its exclusive decays into twenty-six different final states containing charged light--quark hadrons, pion, kaons, and antiprotons.  A similar analysis of $\Upsilon(1S)$ decays is made, and it confirms the identification of $\eta_b(1S)$.

An early attempt to identify $\eta_b(1S,2S)$ in the inclusive allowed M1 radiative decays, $\Upsilon(1S,2S)\to\gamma\eta_b(1S,2S)$ by detecting the low energy ($<100$~MeV) transition photons was unsuccessful~\cite{etaballowincl}.   Successful identification of $\eta_b(1S)$ by BaBar~\cite{etab}, and its confirmation by CLEO~\cite{etab}, was only made possible by detecting the $\sim920$~MeV and $\sim610$~MeV transitions photons in the ``forbidden'' M1 decays $\Upsilon(3S,2S)\to\gamma\eta_b(1S)$.  These M1 transitions have zero overlap between the initial and final states in the lowest order. They become finite only because of relativistic and higher--order effects, and theoretical predictions for them are notoriously difficult and unreliable.  For example, a recent calculation in the framework of non-relativistic effective field theory of QCD (pNRQCD)~\cite{BJV06} predicted a branching fraction for the decay $\Upsilon(2S)\to\gamma\eta_b(1S)$ more than two orders of magnitude larger than that measured by BaBar~\cite{etab}.  In contrast, ``allowed'' M1 transitions, $\Upsilon(nS)\to\gamma\eta_b(nS)$ are relatively simple, with the wave function overlaps between the initial and final states being essentially unity.  They are therefore very attractive from the theoretical point of view.  Since the inclusive radiative transitions are essentially impossible to measure, the only hope is to identify $\eta_b(nS)$ in radiative decays of $\Upsilon(nS)$ by ``tagging'' $\eta_b(nS)$ by their exclusive hadronic decays.

In this paper, we report on the study of the reaction
\begin{multline*}
\Upsilon(2S)\to\gamma\eta_b(2S),~\eta_b(2S)\to X, \\
(X=4,6,8,10~\pi^\pm,K^\pm,p/\bar{p})
\end{multline*}
We also report on an identical study of $\Upsilon(1S)\to\gamma\eta_b(1S),~\eta_b(1S)\to X$, which provides a useful check of our analysis procedure.
We use data obtained with the CLEO~III detector at the Cornell Electron Storage Ring, CESR.  The data consist of $(9.32\pm 0.19)\times10^6~\Upsilon(2S)$ and $(20.82\pm 0.41)\times10^6~\Upsilon(1S)$.  To develop event selection criteria we use Monte Carlo (MC) simulations and data consisting of $(5.88\pm 0.12)\times10^6~\Upsilon(3S)$. 

The CLEO III detector, which has been described before \cite{cleodetector}, consists of a CsI electromagnetic calorimeter, an inner silicon vertex detector, a central drift chamber, and a ring-imaging Cherenkov (RICH) detector, all inside a superconducting solenoid magnet with a 1.5 T magnetic field. The detector has a total acceptance of 93$\%$ of $4\pi$ for charged and neutral particles.  The photon energy resolution in the central ($83\%$ of $4\pi$) part of the calorimeter is about $2\%$ at $E_{\gamma}=1~\gev$ and about $5\%$ at $100\;\,\mev$.  The charged particle momentum resolution is about 0.6$\%$ at $1~\gevc$.

We select events that have 4, 6, 8, or 10 charged particle tracks with zero net charge  and at least one photon candidate.  The charged tracks are required to be well-measured and consistent with coming from the interaction point.  The photon candidates are calorimeter showers which lie within the ``good barrel'' or ``good endcap'' regions, $|\cos\theta|<0.81$ and $|\cos\theta|=0.85-0.93$, respectively, where $\theta$ is the polar angle with respect to the incoming positron direction.  They are required to contain at least 10~MeV of energy, to not contain any of the few known noisy calorimeter cells, and to have a transverse energy distribution consistent with an electromagnetic shower.   
Analysis of simulated events shows that the largest background to the low energy transition photons comes from calorimeter showers due to the interaction of the final state charged hadrons with detector elements.  To reduce this background, we make the isolation requirement that a photon candidate  must be separated from the nearest charged track by 50~cm.  
To ensure that there is no contribution from photons from $\pi^0\to\gamma\gamma$ decays, we reject photon candidates which make a two-photon invariant mass of $M(\gamma\gamma)=M(\pi^0)\pm25$~MeV with any other photon candidate in the event. 

Charged tracks are identified as $\pi^\pm$, $K^\pm$, and $p/\bar{p}$ using $dE/dx$, the energy loss in the drift chamber, and information from the RICH subdetector. 
To utilize $dE/dx$ information, for each particle hypothesis, $X=\pi,~K,~p$ or $\bar{p}$, we calculate $\chi_{X}^{dE/dx}=[(dE/dx)_\mathrm{measured}-(dE/dx)_\mathrm{predicted}]/\sigma_{X}$,  for hypothesis $X$, and $\sigma_{X}$ is the standard deviation of the measured $dE/dx$ for hypothesis $X$.
For higher momentum tracks with $|\cos\theta|<0.8$, we use the combined log-likelihood variable
$$\Delta \mathcal{L}_{X,Y} = (\chi_X^{dE/dx})^2 - (\chi_Y^{dE/dx})^2 + L_X^{RICH} - 
L_Y^{RICH}$$
where $L_{X}^{RICH}$ are the log--likelihoods for a particular hypothesis obtained from measurement in the RICH subdetector.

For low momentum tracks with $p<0.6$~GeV for $\pi^\pm$ and $K^\pm$, and $p<1.5$~GeV for $p/\bar{p}$, we only use $dE/dx$ information, and require the measured $dE/dx$ to be within $3\sigma$ of the expected energy loss for the particle hypothesis.  We also require $|\chi_\pi^{dE/dx}| < |\chi_K^{dE/dx}|$ for charged pions, $|\chi_K^{dE/dx}| < |\chi_\pi^{dE/dx}|$ for charged kaons, and $|\chi_p^{dE/dx}| < |\chi_K^{dE/dx}|$ for $p/\bar{p}$.  For higher momentum tracks, we require $\Delta\mathcal{L}_{\pi,K}<0$ for charged pions, $\Delta\mathcal{L}_{K,\pi}<0$ for charged kaons, and $\Delta\mathcal{L}_{p,K}<0$ for $p/\bar{p}$.  We remove contamination of electrons by rejecting events with tracks which have a ratio between the energy deposited in the calorimeter and the momentum measured in the drift chamber, $E/p=0.9-1.1$. 
 We reconstruct $K_S^0\to\pi^+\pi^-$ by requiring these decays to have a common vertex that is displaced $>3\sigma$ from the interaction point.

We reconstruct the $\eta_b(nS)$ candidate in the following 26 decay modes: $2(\pi^+\pi^-)$, $3(\pi^+\pi^-)$, $4(\pi^+\pi^-)$, $5(\pi^+\pi^-)$,
$K^+K^-\pi^+\pi^-$, $K^+K^-2(\pi^+\pi^-)$, $K^+K^-3(\pi^+\pi^-)$, $K^+K^-4(\pi^+\pi^-)$,
$2(K^+K^-)$, $2(K^+K^-)\pi^+\pi^-$, $2(K^+K^-)2(\pi^+\pi^-)$, $2(K^+K^-)3(\pi^+\pi^-)$, 
$p\bar{p}\pi^+\pi^-$, $p\bar{p}2(\pi^+\pi^-)$, $p\bar{p}3(\pi^+\pi^-)$, $p\bar{p}4(\pi^+\pi^-)$,
$p\bar{p}K^+K^-\pi^+\pi^-$, $p\bar{p}K^+K^-2(\pi^+\pi^-)$, $p\bar{p}K^+K^-3(\pi^+\pi^-)$, 
$K_S^0K^\pm\pi^\mp$, $K_S^0K^\pm\pi^\mp\pi^+\pi^-$, $K_S^0K^\pm\pi^\mp2(\pi^+\pi^-)$, $K_S^0K^\pm\pi^\mp3(\pi^+\pi^-)$, 
$2K_S^0\pi^+\pi^-$, $2K_S^02(\pi^+\pi^-)$, $2K_S^03(\pi^+\pi^-)$.

To select events with well--measured hadrons, we fit the reconstructed hadrons to a common vertex, and require that the reduced $\chi^2$ of the fit is $\chi^2/d.o.f.<4$.

To reconstruct the full event including both the hadrons and the transition photon, we perform a 4C kinematic fit constraining the combination of a photon candidate and the hadronic final state to have the center-of-mass four-momentum of zero (except for a small contribution due to the finite beam crossing angle).  We perform this fit for each signal photon candidate in the event and pick the fit with the lowest $\chi^2$.  Henceforth, we use the constrained hadronic mass which has the much better mass resolution ($\sigma\approx5$~MeV) of the photon.  To reject $\Upsilon(nS)\to\mathrm{hadrons}$ events combined with a fake photon, we use the reduced $\chi^2$ of the 4C fit. We require $\chi^2/d.o.f.<4$.

As noted by BaBar, because there is no preferred direction in the decay of the spin--zero $\eta_b$, there is weak correlation between the signal photon momentum in the center-of-mass frame with the thrust axis calculated for the hadrons from the decay of $\eta_b$.  In contrast, the same correlation is strong for the background events.  Therefore, the signal-to-background ratio varies with the angle $\theta_T$ between the photon and the thrust vector, and a cut on $|\cos\theta_T|$ is very useful in rejecting background.  This was confirmed by both BaBar and CLEO in their identification of $\eta_b(1S)$, and is found to be also true in MC simulations in the present case. 
A cut to accept events with $|\cos\theta_T|<0.5$ is found to be optimum.

The efficiencies for individual decay modes were determined by MC simulations.  They range from 8.2\% to 0.5\%, depending on event multiplicity.

\begin{figure}
\begin{center}
\includegraphics[width=3.5in]{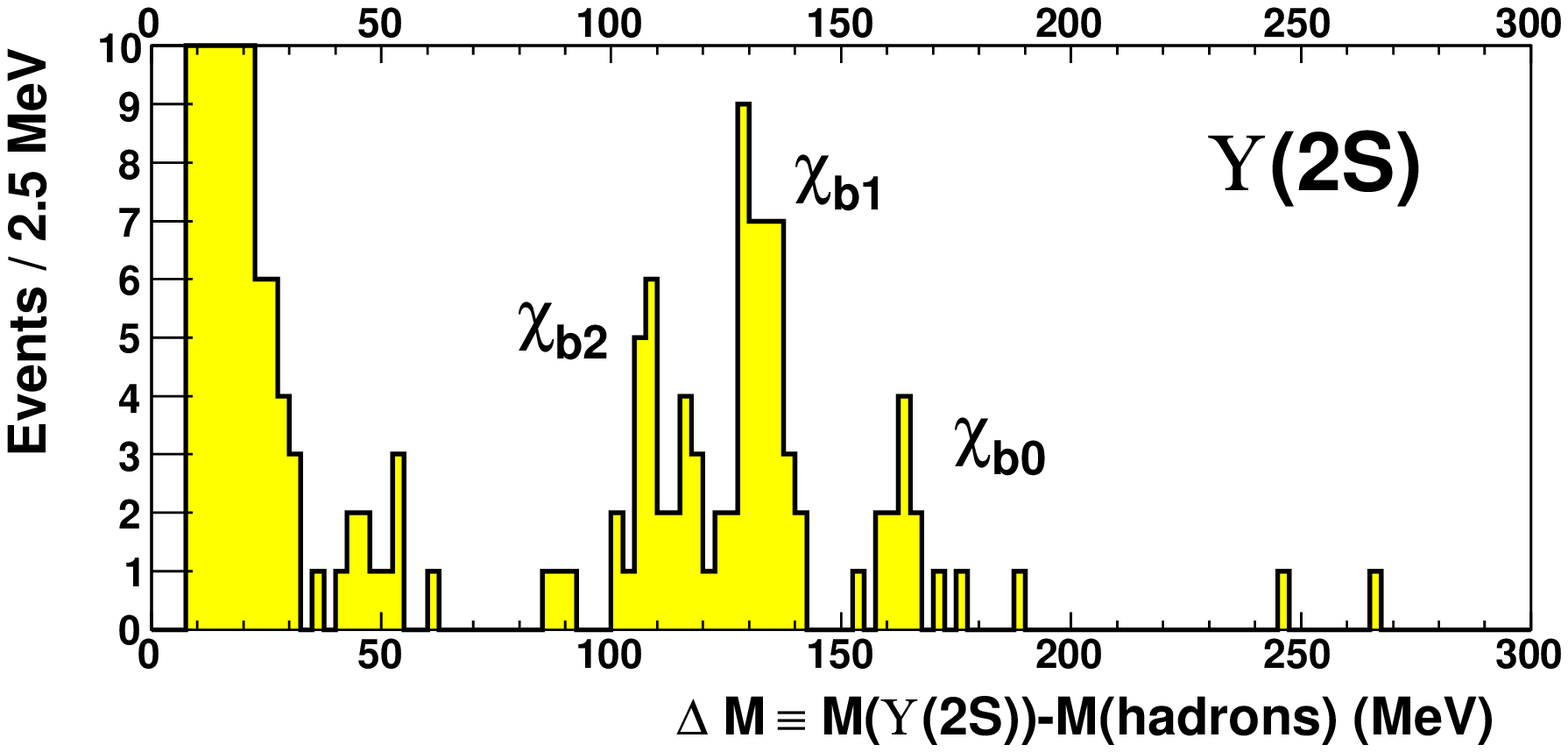}
\includegraphics[width=3.5in]{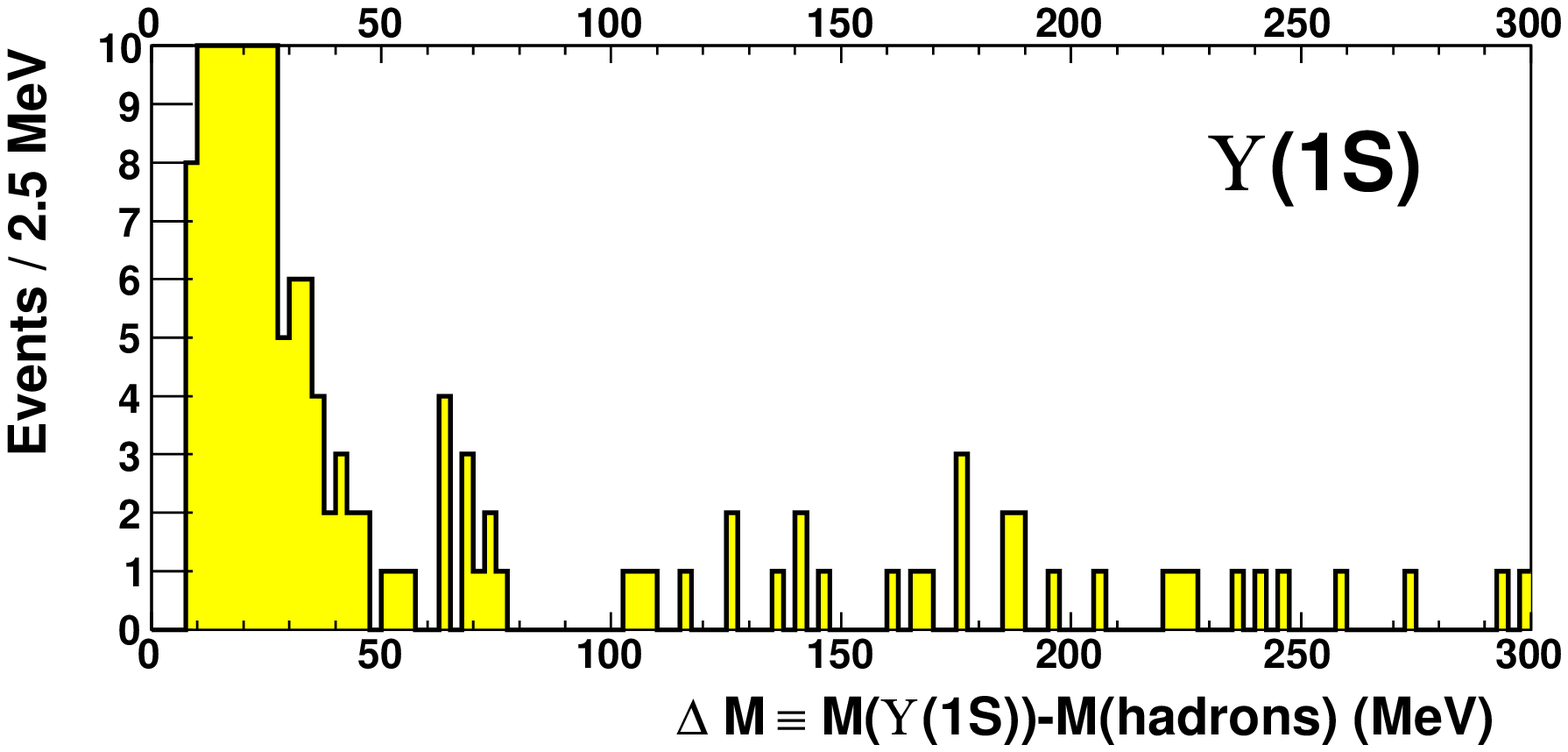}
\end{center}

\caption{Invariant mass distributions  in terms of $\Delta M \equiv M(\Upsilon(2S,1S))-M(\mathrm{hadrons})$ for (top) $\Upsilon(2S)$ and (bottom) $\Upsilon(1S)$ data with final event selections.}

\label{fig:spect}
\end{figure}

\begin{figure*}
\begin{center}
\includegraphics[width=3.1in]{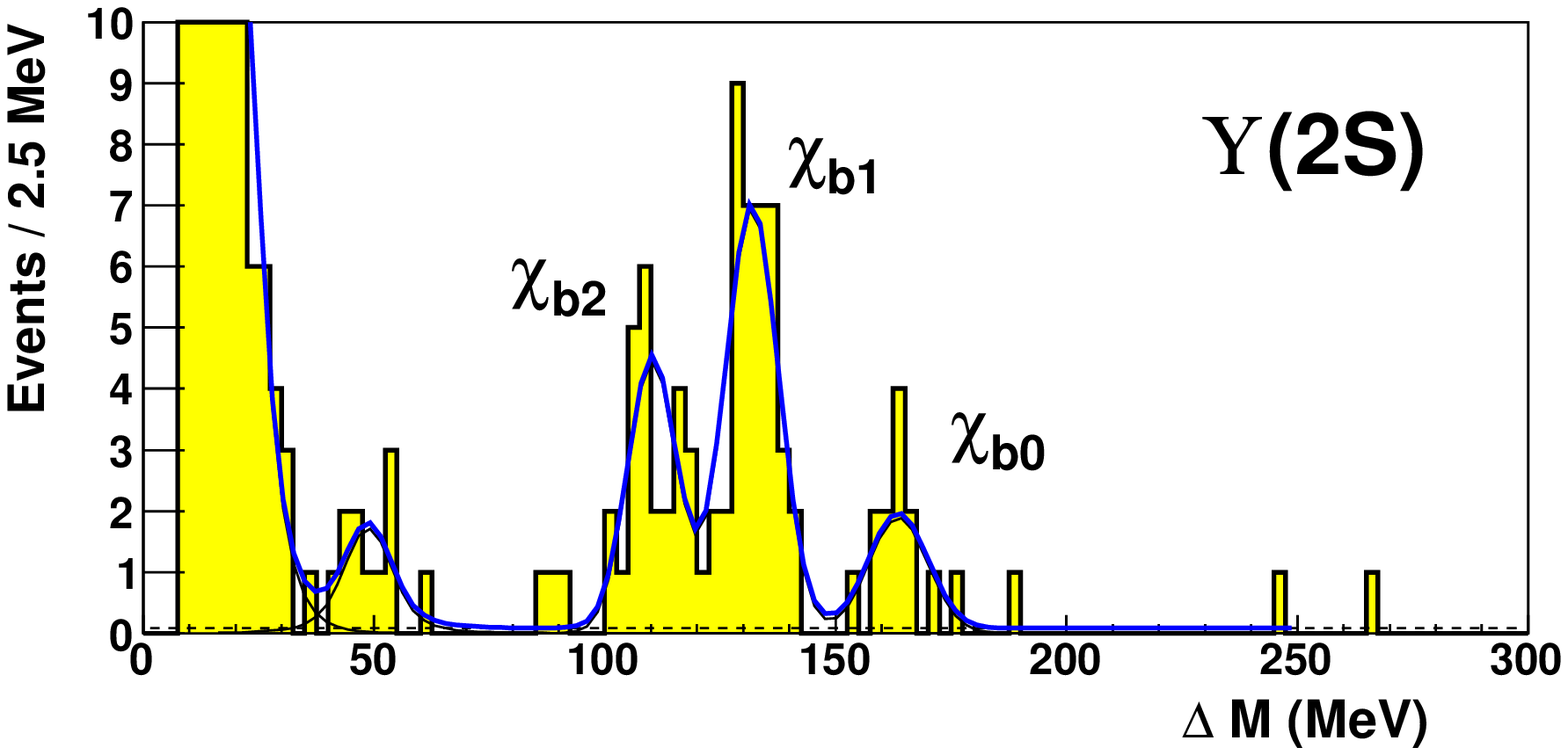}
\includegraphics[width=3.1in]{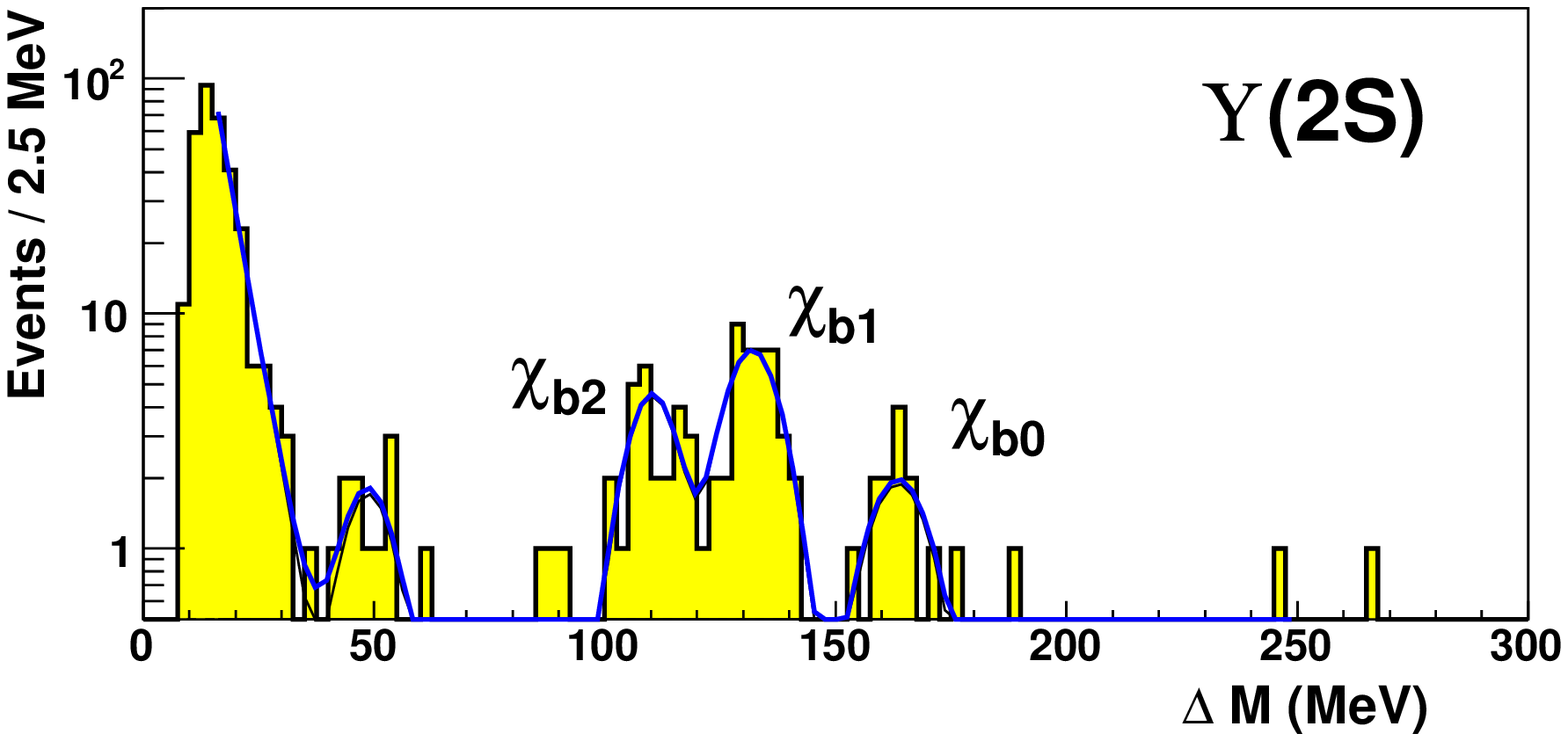}

\includegraphics[width=3.1in]{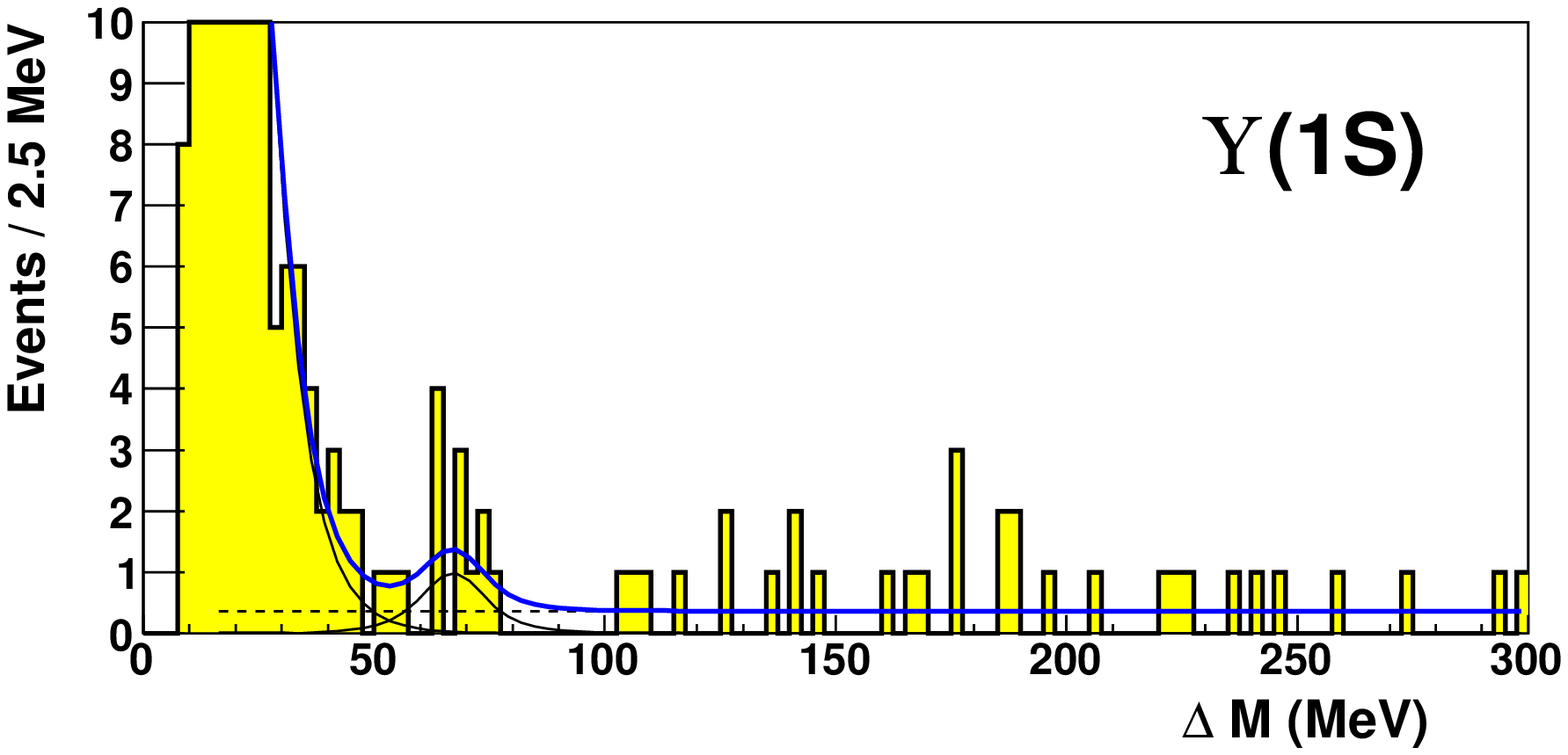}
\includegraphics[width=3.1in]{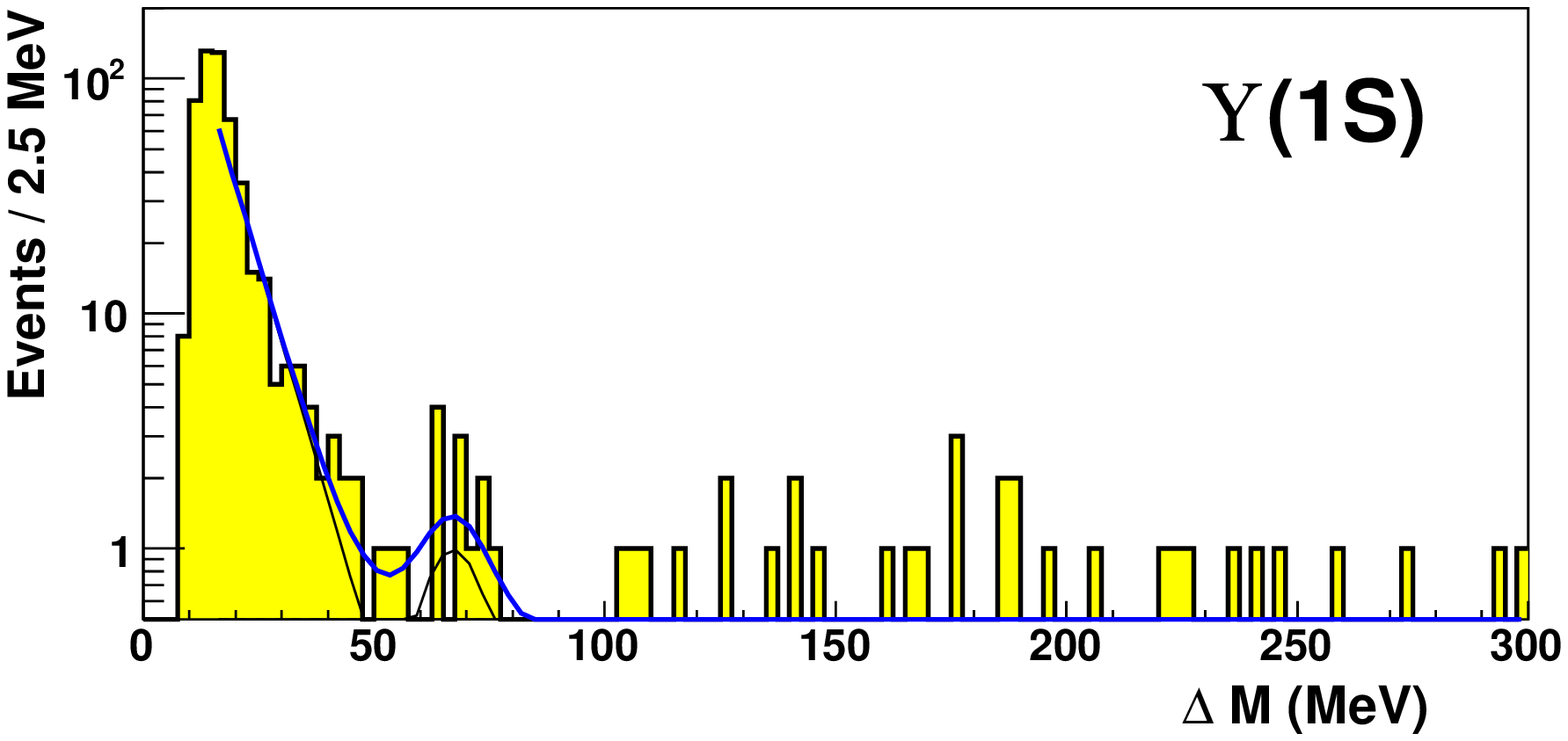}

\end{center}
\caption{Distributions of $\Delta M \equiv M(\Upsilon(2S,1S))-M(\mathrm{hadrons})$: (top row) $\Upsilon(2S)$ data, (bottom row) $\Upsilon(1S)$ data are shown with both (left column) linear and (right column) logarithmic scales.  The best fit curves are shown as the thick solid curve and the individual components are shown as the dashed and thin solid curves.}

\label{fig:finalfits}
\end{figure*}

\begin{table*}
\caption{Results of fits to the $\Delta M(2S,1S)$ data distributions, as described in the text.  Only statistical uncertainties are given.  The product branching fraction is $\mathcal{B}_1\times\mathcal{B}_2\equiv\mathcal{B}_1(\Upsilon(nS)\to\gamma\eta_b(nS))\times\sum^{26}_{i=1} \mathcal{B}_{2i}(\eta_b(nS)\to h_i)$.  The first errors are statistical, and the second errors are systematic, as detailed in Table~\ref{tbl:systs}.}

\setlength{\tabcolsep}{8pt}

\begin{center}
\begin{tabular}{lcccccc}
\hline \hline
  & $N$ & $\Delta M_\mathrm{hf}$ (MeV) & $M$ (MeV) & $\chi^2/d.o.f.$   & signif. ($\sigma$) &  $\mathcal{B}_1\times\mathcal{B}_2\times10^6$  \\
\hline
$\eta_b(2S)$ & $11.4^{+4.3}_{-3.5}$ & $48.7\pm2.3\pm2.1$ & $9974.6\pm2.3\pm2.1$ & $91.8/103$  & $4.9$   & $46.2^{+29.7}_{-14.2}\pm10.6$ \\
$\eta_b(1S)$ & $10.3^{+4.9}_{-4.1}$ & $67.1\pm3.4\pm2.3$ & $9393.2\pm3.4\pm2.3$ & $114.6/107$  & $3.1$   & $30.1^{+33.5}_{-7.4}\pm7.5$  \\

\hline \hline
\end{tabular}
\end{center}

\label{tbl:finalfitresults}

\end{table*}

The  invariant mass distributions for $\Upsilon(1S)$ and $\Upsilon(2S)$ data with final event selections are shown in Fig.~\ref{fig:spect} in terms of $\Delta M\equiv M(\Upsilon(nS))-M(\mathrm{hadrons})$.
At the smallest values of $\Delta M$, the distributions show the large, steeply falling contributions due to $\Upsilon(nS)$ decays.  
Enhancements are seen at $\Delta M\approx50$~MeV in the $\Upsilon(2S)$ data and at $\Delta M\approx70$~MeV in the $\Upsilon(1S)$ data.  These appear to be well separated from the $\Upsilon$ contributions.  
In the $\Upsilon(2S)$ data, the contributions from $\Upsilon(2S)\to\gamma\chi_{bJ}(1P)$, $\chi_{bJ}(1P)\to\mathrm{hadrons}$ are also seen in the range $\Delta M\approx100-200$~MeV.
The distributions in  Fig.~\ref{fig:spect} have been fitted as follows. 

The non--peaking backgrounds due to continuum and misidentifications in the $\Upsilon(2S)$ data is essentially zero both below and above the $\chi_{bJ}(1P)$ peaks, and in the $\Upsilon(1S)$ data it is nearly constant ($\sim0.4~\mathrm{count}/2.5$~MeV~bin) in the region $\Delta M=100-300$~MeV.

The choice of the fit function for the large, rapidly falling $\Upsilon(2S)$ and $\Upsilon(1S)$ contributions at small $\Delta M$ is important.
It is not possible to obtain the total shapes of the $\Upsilon(2S)$ and $\Upsilon(1S)$ contributions from MC simulations.  We have made a large number of MC simulations, and find that the predicted shapes of these contributions differ substantially between different decays and multiplicities, particularly in the tail regions.  Further, they can not be added to produce the composite shape because the relative proportions of the individual contributions are not known.
Hence, an empirical approach to fit it was adopted, and fits with different fit functions (exponentials of the form $\exp( ax + bx^2 + cx^3 + \cdots )$) were tried.  The best fits were consistently obtained with a single exponential.  Single exponentials were also found to best fit the data for $\Upsilon(3S)$.
In the left panels of Fig.~\ref{fig:finalfits} we show the fits in linear plots. In the right panels we show the same fits in log plots to illustrate that the single exponentials fit the $\Upsilon(nS)$ contributions very well,  and the enhancements at $\sim70$~MeV and $\sim50$~MeV in the $\Upsilon(1S)$ and $\Upsilon(2S)$ data received very little contribution from the ``tails'' of $\Upsilon(1S,2S)$.  

We fit the peaks in the $\Delta M$ distributions with Breit--Wigner shapes convolved with the known Gaussian experimental resolution functions which have widths which vary from $\sigma=4.2$~MeV at $\Delta M=50$~MeV to $\sigma=6.4$~MeV at $\Delta M=165$~MeV.  The Breit--Wigner width of the enhancement in $\Upsilon(2S)$ at $\Delta M\approx50$~MeV is assumed to be 5~MeV.  The Breit--Wigner width of the enhancement in $\Upsilon(1S)$ at $\Delta M\approx70$~MeV, attributed to $\eta_b(1S)$, is assumed to be 10~MeV.  The $\chi_{bJ}(1P)$ peaks are fitted with the Gaussian resolution widths.  
The masses of the $\chi_{bJ}(1P)$ peaks are found to be in agreement with their known masses within $1.1\pm0.8$~MeV on average.

The fit results are listed in Table~\ref{tbl:finalfitresults}.

The significance values in Table~\ref{tbl:finalfitresults} are determined as $\sigma\equiv\sqrt{-2\ln(L_0/L_{max})}$, where $L_{max}$ is the maximum likelihood returned by the fits including the enhancements at $\sim70$~MeV and $\sim50$~MeV, and $L_0$ is the likelihood returned by fits without these enhancements.

The fitted value of the $\sim70$~MeV enhancement in the $\Upsilon(1S)$ data is $\Delta M = 67.1\pm3.4(\mathrm{stat})$~MeV, and the observation has a significance of $3.1\sigma$.  It is naturally identified as being due to $\eta_b(1S)$, and leads to $\Delta M_\mathrm{hf}(1S)_{b\bar{b}}=67.1\pm3.4(\mathrm{stat})$~MeV, in good agreement with the PDG average of $69.3\pm2.8$~MeV~\cite{pdg}.  The fitted value of the $\sim50$~MeV enhancement in the $\Upsilon(2S)$ data is $48.7\pm2.3(\mathrm{stat})$~MeV, and the observation has a significance of $4.9\sigma$.  We can not find any explanation for this $4.9\sigma$ enhancement except to attribute it to $\eta_b(2S)$.  Henceforth, we refer to it as such.  Thus, we determine $\Delta M_\mathrm{hf}(2S)_{b\bar{b}}=48.7\pm2.3(\mathrm{stat})$~MeV.  This constitutes the first determination of hyperfine splitting in the bottomonium $2S$ radial excitation.

To confirm the likelihood determination of the significance of the $\eta_b(2S)$ enhancement, we have made a MC determination (taking systematic uncertainties into account) of the probability that the observed enhancement can arise due to a statistical fluctuation anywhere in the range $\Delta M=35-70$~MeV.  In $10^9$ trials, 4565 such fluctuations were found.  This corresponds to a significance of $4.6\sigma$ for the observed enhancement.  

We calculate product branching fractions corresponding to the observed counts $N_i(\mathrm{obs})$ in individual hadronic decay modes $h_i$ using MC--determined efficiencies $\epsilon_i$ as 
\begin{multline}
\mathcal{B}_1[\Upsilon(nS)\to\gamma\eta_b(nS),\gamma\chi_{bJ}(1P)] \\ \times\mathcal{B}_{2i}[(\eta_b,\chi_{bJ})\to h_i] = \frac{N_i(\mathrm{obs})}{\epsilon_i\times N(\Upsilon(nS))},
\end{multline}
where $N(\Upsilon(nS))$ refers to the number of $\Upsilon(nS)$ in the data samples.  Because of the very small number of counts in individual decays ($\le4$ in $\eta_b$ and individual $\chi_{bJ}$ transitions) statistically significant results for $\eta_b$ and $\chi_{bJ}$ transitions can only be obtained by summing over all the decay channels to obtain $\sum_i \mathcal{B}_1\mathcal{B}_{2i}$.  For the $\eta_b$, these are listed in Table~\ref{tbl:finalfitresults}.  Admittedly, these are still rather crude results, and statistical errors do not represent the uncertainties reliably.  A better measure of the uncertainty is provided by comparing the sum $\sum_i \mathcal{B}_1\mathcal{B}_{2i}$ over the three $\chi_{bJ}$ states and the 5 decays measured by us for each ($\to3(\pi^+\pi^-)$, $4(\pi^+\pi^-)$, $K^+K^-2(\pi^+\pi^-)$, $K^+K^-3(\pi^+\pi^-)$, $K_S^0K^\pm\pi^\mp\pi^+\pi^-$) with the sum of the published results for them~\cite{chib-hadronic}.  Our results, based on different event selections and much smaller statistics, agree with the published results within a factor 1.5.

\begin{table}[!tb]
\caption{Summary of systematic uncertainties and their sums in quadrature.}
\begin{center}
\begin{tabular}{l|cc|cc}
\hline \hline
Sources of    
  & \multicolumn{2}{c|}{$\Upsilon(1S)$ Results}  & \multicolumn{2}{c}{$\Upsilon(2S)$ Results} \\
\cline{2-5}
systematic uncertainties  & $\Delta M_\mathrm{hf}$  & $\mathcal{B}_1\!\times\!\mathcal{B}_2$   & $\Delta M_\mathrm{hf}$  & $\mathcal{B}_1\!\times\!\mathcal{B}_2$  \\
(ranges of their variations)  & (MeV) & (\%)  & (MeV) & (\%) \\
\hline
Number of $\Upsilon(nS)$ &    ---    &  $\pm2$  &    ---    & $\pm2$ \\
Mass Calibration (from $\chi_{bJ}$)  & $\pm2.0$  &   ---    & $\pm2.0$  & --- \\
Reconstruction \& PID    &    ---    & $\pm16$  &    ---    & $\pm16$ \\
Detector Resolution ($\pm10\%$)     & $\pm0.2$  &  $\pm1$  & $\pm0.1$  & $\pm5$ \\
Fit Range  
  & $\pm0.8$  & $\pm14$  & $\pm0.3$  & $\pm7$ \\
\hfill $12.5\!-\!20\to250\!-\!300~\mathrm{MeV}$ & & & & \\ 
$\Gamma(\eta_b(1S))\!=\!5\!-\!15$~MeV  & $\pm0.8$  & $\pm11$  & $\pm0.1$  & $\pm12$ \\
$\Gamma(\eta_b(2S))\!=\!2.5\!-\!7.5$~MeV  & & & & \\
$\Upsilon(nS)$, 1st$\to$3rd order expon. & $\pm0.2$ & $\pm7$ & $\pm0.2$ & $\pm5$ \\
Bin Size ($1\!-\!2.5$~MeV)   & $\pm0.1$  &  $\pm1$  & $\pm0.4$  & $\pm4$ \\
\hline
Total & $\pm2.3$ & $\pm25$ & $\pm2.1$ & $\pm23$ \\
\hline\hline
\end{tabular}
\end{center}
\label{tbl:systs}
\end{table}

Based on the fitted masses of the $\chi_{bJ}(1P)$ resonances, we conservatively assign the systematic uncertainty of $\pm2.0$~MeV in our mass calibration.
The systematic uncertainties in our results due to other possible sources were taken to be equal to the maximum variations in $\Delta M_\mathrm{hf}$ and $\mathcal{B}_1\times\mathcal{B}_2$ found in varying the source parameters from their nominal values.
The range of variations for different parameters and the maximum variations observed in $\Delta M_\mathrm{hf}$ and $\mathcal{B}_1\times\mathcal{B}_2$ are listed in Table~\ref{tbl:systs}.  The uncertainty in number of $\Upsilon(nS)$ produced is estimated to be $2\%$.  Event reconstruction and PID uncertainties vary for the different decay modes.  The maximum uncertainties correspond to the largest multiplicity decays.  To be conservative, in Table~\ref{tbl:systs} we assign these maximum uncertainties, 16\% for $\eta_b(1S)$ and  $\eta_b(2S)$, also to the sum of all modes.  
The systematic uncertainties added in quadrature are $\pm2.3$~MeV in $\Delta M_\mathrm{hf}(1S)$ and $\pm2.1$~MeV in $\Delta M_\mathrm{hf}(2S)$, and $\pm25\%$ and $\pm23\%$ in the corresponding $\mathcal{B}_1\times\mathcal{B}_2$.  Our final results are obtained by adding these in quadrature to the statistical errors in Table~\ref{tbl:finalfitresults}.

Our results for the 1S state, $M(\eta_b(1S)) = 9393.2\pm4.1~\mathrm{MeV}$ and $\Delta M_\mathrm{hf}(1S)_{b\bar{b}} = 67.1\pm4.1~\mathrm{MeV}$, agree with previous determinations~\cite{etab}.  Our identification of the $\eta_b(2S)$ state at a $\sim5\sigma$ level leads to
\begin{align*}
M(\eta_b(2S)) & = 9974.6\pm3.1~\mathrm{MeV} \\
\Delta M_\mathrm{hf}(2S)_{b\bar{b}} & = 48.7\pm3.1~\mathrm{MeV} %\\
\end{align*}

In summary, we have presented evidence for the first successful observation of $\eta_b(2S)$,  and the hyperfine splitting of the bottomonium 2S state.
Unquenched lattice predictions for radial excitations are admittedly not yet very reliable~\cite{ukqcd,fnallat,meinel}.  The latest of these calculations~\cite{hpqcd} by the HPQCD~Collaboration obtains $\Delta M_\mathrm{hf}(1S)_{b\bar{b}}=70\pm9$~MeV, in good agreement with its experimental value, and predicts $\Delta M_\mathrm{hf}(2S)_{b\bar{b}}=35\pm3$~MeV~\cite{hpqcd}. 

This investigation was done using CLEO data, and as members of the former CLEO Collaboration we thank it for this privilege.  
We wish to thank H.~Vogel for useful comments.
This research was supported by the U.S. Department of Energy.

\end{document}